\documentstyle[floats,epsfig,aps,prb]{revtex}

\newcommand{\dfrac}[2]{\frac{\strut \displaystyle{#1}}{\displaystyle{#2}}}
\def\bfk{\mbox{\boldmath $k$}}

\def\bfA{\mbox{\boldmath $A$}}
\def\bfe{\mbox{\boldmath $e$}}
\def\bfr{\mbox{\boldmath $r$}}

\begin{document}

\draft

\twocolumn[\hsize\textwidth\columnwidth\hsize\csname 
@twocolumnfalse\endcsname

\title{\large\bf Magnetization plateau
  in a two-dimensional multiple-spin exchange model}

\author{Tsutomu Momoi, Harumi Sakamoto and Kenn Kubo}
\address{Institute of Physics, University of Tsukuba, 
Tsukuba, Ibaraki 305-8571, Japan}
\date{6 Oct. 1998, revised 14 Dec. 1998}

\maketitle

\medskip
\begin{abstract}
We study a multiple-spin exchange model on a
triangular lattice, which is a possible model for low-density solid {}$^3$He
films. Due to strong competitions between ferromagnetic three-spin
exchange and antiferromagnetic four-spin one, 
the ground states are highly degenerate in the classical limit. 
At least $2^{L/2}$-fold degeneracy exists on the $L\times L$
triangular lattice except for the $SO(3)$ symmetry. 
In the magnetization process, we found a plateau at
$m/m_{\rm sat}=1/2$, in which the ground state is {\it uuud} state 
(a collinear state with four sublattices). 
The 1/2-plateau appears due to the strong four-spin exchange interaction. 
This plateau survives against both quantum and thermal fluctuations. 
Under a magnetic field which realizes the {\it uuud} ordered state, 
a phase transition occurs at a finite temperature due to 
breakdown of translational symmetry. 
We predict that low-density solid {}$^3$He thin films may 
show the $1/2$-plateau in the magnetization process. 
Experimental observation of the plateau will verify strength 
of the four-spin exchange. 
It is also discussed that this magnetization
plateau can be understood as an insulating-conducting transition in 
particle picture.  
\end{abstract}
\vskip0.3pc]
\narrowtext

\section{Introduction}
\indent
In localized fermion systems, the magnetic interaction 
comes from permutations of particles.\cite{Dirac,Thouless}
Multiple-spin exchanges, e.g.\ cyclic exchanges of three or four spins, 
have been revealed to be strong 
in nuclear magnetism of two-dimensional (2D) solid {}$^3$He films. 
Many experimental\cite{SiqueiraNCSa,RogerBBCG,IshidaMYF} and
theoretical\cite{DelrieuRH,RogerWKB,BernuCL92,KatanoH,MisguichBLC}
studies suggested that exchange interactions of more-than-two spins 
are dominant in this system. A ferromagnetic behavior changes 
to antiferromagnetic one when the coverage of {}$^3$He 
decreases.\cite{Godfrin,SchifferOOF,SiqueiraLCS93,MorishitaIYF,%
SiqueiraNCSb}  
This tendency can be understood in terms of multiple-spin 
exchange (MSE): in fully packed systems, the three-spin exchange is
dominant\cite{DelrieuRH} and it is ferromagnetic (as shown by
Thouless\cite{Thouless}), and, in loosely packed systems, the four- and 
six-spin exchanges become strong and favor 
antiferromagnetism. Effects of multiple-spin exchanges are not yet
fully understood especially for the low-density
region.\cite{Greywall,IshidaMYF} For example, 
recent specific-heat data at low densities show 
a peculiar behavior, which have a double-peak 
structure, and they also show that the ground state seems
to be spin liquid (disordered) and the spin excitation gap is
vanishing (or quite small).\cite{IshidaMYF} 

The multiple-spin exchange (MSE) model has been studied to describe 
the nuclear magnetism of three-dimensional solid
{}$^3$He.\cite{RogerHD} A general form of
the spin Hamiltonian of quantum solid is\cite{Dirac,Thouless} 
${\cal H}=-\sum_n (-1)^n J_n \sum_{P_n} P_n$, where $P_n$ and $J_n(\le 
0)$ denote cyclic
permutation of $n$ spins and its exchange constant, respectively. 
For the 2D system, recent theoretical 
calculations\cite{BernuCL92,KatanoH,MisguichBLC} and experimental
measurements\cite{RogerBBCG} found that the exchange frequencies
satisfy $|J_3|>|J_2|>|J_4| \gtrsim |J_6| \gtrsim |J_5|$ 
on the triangular lattice. 
In this paper, we consider a spin model with the two-, three-,
and four-spin exchanges on the triangular lattice, which is the
simplest 2D MSE model. Since the
three-spin exchange can be transformed to the two-spin ones, the
Hamiltonian can be written with two parameters $J$ and $K$ as
\begin{equation}\label{Hamiltonian}
{\cal H}= J \sum_{\langle i,j \rangle} 
        \mbox{\boldmath $\sigma$}_i \cdot \mbox{\boldmath $\sigma$}_j 
  + K \sum_p h_p - \mu B\sum_i \sigma_i^z,
\end{equation}
where $\mbox{\boldmath $\sigma$}_i$ denote Pauli matrices. 
The last term means the Zeeman energy, where $\mu$ denotes the
nuclear magnetic moment of {}$^3$He and $B$ the magnetic field. 
The parameter $J(=J_3-J_2/2)$ is negative for most of densities, 
(but it can change the sign,) and $K(=-J_4/4)$ is always positive 
($K\ge 0$). The first and the second summations run over all pairs of
nearest neighbors and all minimum diamond clusters, respectively. 
The explicit form of $h_p$ for four sites $(1,2,3,4)$ is 
\begin{eqnarray}
 h_p &=& 4(P_4 + P_4^{-1})-1 \nonumber\\
     &=& \sum_{1\le i<j \le 4} 
            \mbox{\boldmath $\sigma$}_i \cdot \mbox{\boldmath $\sigma$}_j 
      + (\mbox{\boldmath $\sigma$}_1 \cdot \mbox{\boldmath $\sigma$}_2)
        (\mbox{\boldmath $\sigma$}_3 \cdot \mbox{\boldmath $\sigma$}_4) 
\nonumber \\
     &+& (\mbox{\boldmath $\sigma$}_1 \cdot \mbox{\boldmath $\sigma$}_4)
         (\mbox{\boldmath $\sigma$}_2 \cdot \mbox{\boldmath $\sigma$}_3) 
      - (\mbox{\boldmath $\sigma$}_1 \cdot \mbox{\boldmath $\sigma$}_3)
        (\mbox{\boldmath $\sigma$}_2 \cdot \mbox{\boldmath $\sigma$}_4),
\end{eqnarray}
where $(1,3)$ and $(2,4)$ are diagonal bonds of the diamond. 
The WKB approximations\cite{RogerWKB,KatanoH} show that exchange
parameters vary 
depending on the particle density: At high densities, the 
exchange $J(\le0)$ is dominant, which mainly originates from the three spin
exchange. As lowering the density, the ratio $|K/J|$ increases
rapidly and hence the four spin exchange $K(\ge 0)$ 
becomes important. This density dependence is consistent with experimental
results of
susceptibility.\cite{Godfrin,SchifferOOF,SiqueiraLCS93,MorishitaIYF,%
SiqueiraNCSb} Since
multiple-spin exchanges produce frustration by themselves and 
strong competitions between exchanges also introduce 
frustration,\cite{Roger90} 
this model is expected to show various complex magnetic behaviors. 

In a previous paper, we studied the ground state of this model in the 
classical limit and found various phases:\cite{KuboM}
(a) For $J<-8K$, the ground state shows the perfect ferromagnetism. 
(b) For $-8K<J<-8K/3$, ground states are highly degenerate. This
degeneracy is a non-trivial one. 
(c) In $-8K/3<J<25K/3$, the ground state has 
a four-sublattice structure with zero magnetization, which we call as 
the tetrahedral structure.\cite{MomoiKN} 
(d)~For $25K/3 <J$, the ground state is the so-called $120^\circ$ structure. 
Thus novel phases (b) and (c) appear due to the four-spin exchange
interactions. In the region (c), we predicted chiral symmetry breaking
at a finite temperature.\cite{MomoiKN} 

The intermediate phase (b) seems to correspond to the parameter region
of 2D low-density solid {}$^3$He. The parameters $J$ and 
$K$ in the low density region are estimated as $J\simeq -1.5$(mK) 
and $K\simeq 0.2$(mK) from susceptibility and specific-heat
data,\cite{RogerBBCG} and $J/K=-4\pm 2$ from path integral Monte Carlo
simulations\cite{BernuCL92,KatanoH,MisguichBLC}. 
The five- and six-spin exchanges are also estimated to be comparable with 
four-spin one. This parameter region almost belongs to
the phase (b) of our study in the classical limit. 
Our previous study\cite{KuboM} shows 
that competitions between the two- and four-spin exchanges are strong in
this phase and many kinds of ground states exist due to
frustration. Furthermore, we found that under the magnetic field a
plateau appears at $m/m_{\rm sat}=1/2$ in the magnetization curve, 
where $m_{\rm sat}$ denotes the saturated magnetization. These 
various unusual phenomena occur due to frustration caused by the multi-spin
exchanges. In this paper we study this phase further, 
especially considering quantum 
effects. Finite-temperature effects are also discussed. 

In section~II we summarize the results for the phase (b) of the 
classical model. Among degenerate ground states, one collinear state, 
which we call $uuud$ state, has the largest magnetization,
$m/m_{\rm sat}=1/2$, and other states show $m/m_{\rm sat}<1/2$. 
In section~III we discuss the quantum model. 
Due to quantum effects, the $uuud$ state disappears from the ground
state at $B=0$ and the ground state belongs to the $S=0$ space. 
Magnetization process of the quantum model is studied in section~IV. 
Under the magnetic field, the $uuud$ ordered state becomes
stable, since it has the largest magnetization, and makes a plateau at
$m/m_{\rm sat}=1/2$ in the magnetization curve. We discuss thermal 
effects in section~V. Under the magnetic field, which realizes the
{\it uuud} ground state, the system shows a finite-temperature phase
transition due to breakdown of translational symmetry. 
Section~VI contains summaries and discussions. 

\section{Classical limit}
\indent
We studied the ground state of the model (\ref{Hamiltonian}) in the
classical limit,\cite{KuboM} 
where the Pauli matrices are replaced to unit vectors
($u^x_i$, $u^y_i$, $u^z_i$) with $|u^x_i|^2 + |u^y_i|^2 + |u^z_i|^2 = 1$. 
We searched the ground state using the mean-field theory and studied 
finite-size systems with Monte Carlo method. 
We searched the ground state restricting ourselves to spin
configurations with up to four-sublattice structures  
within the mean-field theory. To take into account larger
sublattice structures, we studied larger finite-size systems and
searched the minimum energy state with Monte Carlo method, gradually
decreasing temperatures. Here we only discuss the phase (b), where the 
parameters are in $-8K<J<-8K/3$ and competitions of the two- and
four-spin exchanges are strong. 

\begin{figure}[tbp]
  \begin{center}
    \leavevmode
    \epsfig{file=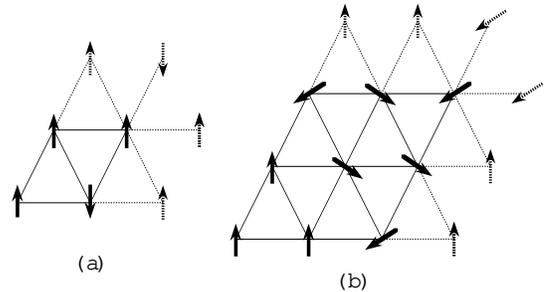,width=2.8in}
  \end{center}
  \caption{Spin configuration of (a) {\it uuud} state with four
    sublattices and (b) a coplanar state with nine sublattices.} 
  \label{fig:classic}
\end{figure}

\begin{figure}[tbp]
  \begin{center}
    \leavevmode
    \epsfig{file=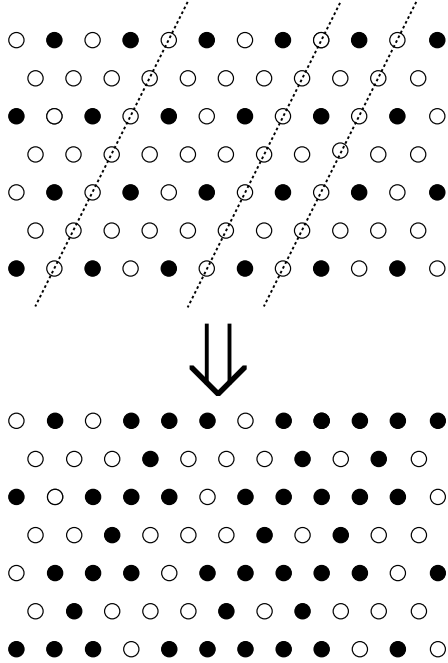,width=2.3in}
  \end{center}    
  \caption{Spin configuration of two kinds of ground states. Up (down)
    spins are denoted by white (black) circles. The upper 
    configuration is that of the {\it uuud} state. The lower one is 
    obtained by reversing 
    the up spins to down on the dashed lines of the upper one.} 
  \label{fig:classic_degen}
\end{figure}

If there is no magnetic field, ground states are highly degenerate for
$-8K<J<-8K/3$. We list some ground states which we found: 
\begin{enumerate}
\item A collinear state with a four-sublattice structure. Up 
  spins are on three sublattices and down spin on the
  other (see Fig.~\ref{fig:classic}a). We call this state 
  {\it uuud} state. 
\item A coplanar state with nine sublattices, whose spin configuration 
  is constructed from three kinds of spin vectors 
  (see Fig.~\ref{fig:classic}b). 
\item Many other ground states can be made out of the {\it uuud} state
  by reversing all up spins to down on some parallel straight lines that 
  consist of only up spins (see Fig.~\ref{fig:classic_degen}). 
\end{enumerate}
All these states have the minimum energy $E/N=-3K$. 
The {\it uuud} state has the largest magnetization $m/m_{\rm
  sat}=1/2$ among these states. The coplanar state shows $S^z=0$ and the 
third series of ground states have a variety of magnetization between 
$-1/2<m/m_{\rm sat}<1/2$. The number of degeneracy except for the 
$SO(3)$ symmetry is at least of order $2^{L/2}$ on the $L\times L$
triangular lattice, which comes from line degrees of freedom for spin
flips. The number of the states in the $S^z=ML$ sector, where $M$ is
an integer in $-L/2\le M \le L/2$, is at least $_{L/2}C_{(L-2M)/4}$ and 
thus the degeneracy is largest in the $S^z=0$ sector. 
This degeneracy does not originate from the symmetry of the Hamiltonian and
instead comes from frustration effects. We suspect that strong frustration 
makes density of states large near the lowest energy 
and hence non-trivial degeneracy appears in the ground states. 

\begin{figure}[tbhp]
  \begin{center}
    \leavevmode
    \epsfig{file=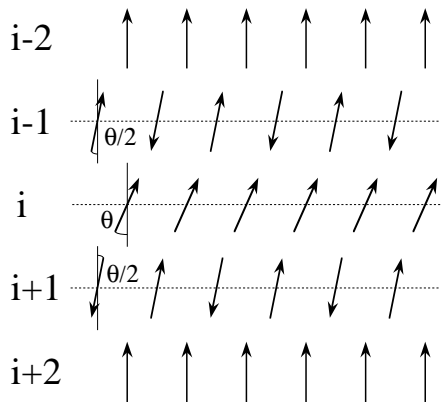,width=2.3in}
  \end{center}    
  \caption{A stripe excitation of the {\it uuud} state. Spins on the 
    $i$th line are rotated with angle $\theta$, and on the
    $i-1$th and $i+1$th lines with angle $\theta/2$. } 
  \label{fig:line_excitation}
\end{figure}

Another characteristic property also appears in excitations. For the
ground states in group 3, a huge number of excitations 
have extremely low energy, 
which is very close to the ground state energy. Consider a
stripe excitation which is shown on Fig.~\ref{fig:line_excitation}. 
On the $i$th
line, rotate spins with angle $\theta$ and, on the $i-1$th and $i+1$th
lines, rotate spins with angle $\theta/2$, where $i$th line contains
only up spins. For small $\theta$, expanding the excitation energy
with $\theta$, one can find that, ${\cal O}(\theta^2)$ terms of the
excitation energy vanish 
and the leading term starts from the order ${\cal O}(\theta^4)$, 
if all spins on both $i-2$th and $i+2$th lines direct
upward. We note that line excitation energy usually depends on the
angle $\theta$ in the quadratic form. 
The condition for the low excitations of this kind to appear is
that all spins on three lines in next neighbors, {\it i.e.}\ on $i-2$th,
$i$th and $i+2$th lines, are in the same direction with each
other. We hence find that these low energy excitations exist for most
of ground states in group 3 and there are a huge number of low-lying
excitations of this kind. 

\begin{figure}[tb]
  \begin{center}
    \leavevmode
    \epsfig{file=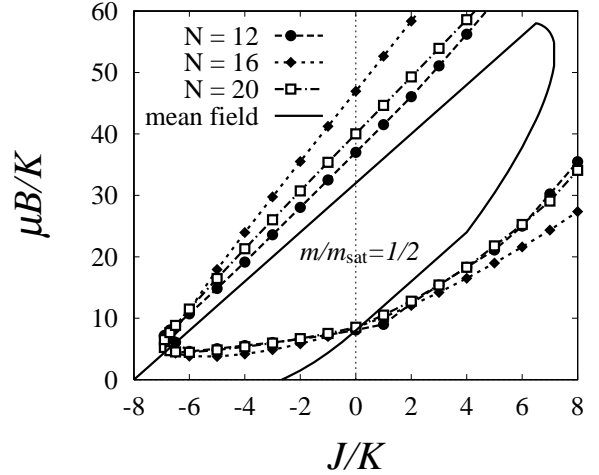,width=3.36in}
  \end{center}    
  \caption{Parameter dependence of phase boundaries of the $m/m_{\rm
      sat}=1/2$ state in the 
    magnetization process at $T=0$, where the 1/2-plateau appears in the
    region surrounded by data. The solid line denotes the result
    from the mean-field theory in the classical limit and
    other data with lines denote the results of the quantum model
    ($S=1/2$) on finite-size systems. } 
  \label{fig:PD}
\end{figure}

\begin{figure}[tbh]
  \begin{center}
    \leavevmode
    \epsfig{file=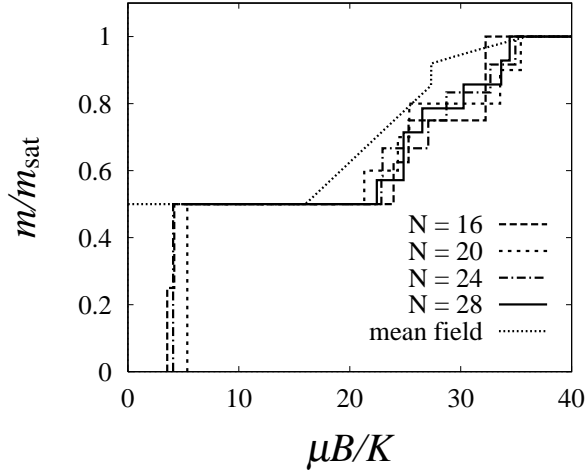,width=3.36in}
  \end{center}
  \caption{Magnetization process at $T=0$ for $J=-4$ and $K=1$. The
    dotted line denotes the result from the mean-field theory 
    in the classical limit\protect\cite{KuboM} and 
    other lines denote the results of the quantum model
    ($S=1/2$) on finite-size systems. }
  \label{fig:mag_pro}
\end{figure}
By applying a weak magnetic field, this degeneracy quickly
disappears. Since the {\it uuud} state has the largest
magnetization among the degenerate states, the {\it uuud} state is
stable under the magnetic field and it has the lowest energy. 
Figure \ref{fig:PD} shows the parameter region where the {\it
  uuud} state becomes the ground state. 
The {\it uuud} state remains to be the ground state up to 
finite magnitude of the magnetic field, which means 
the magnetic susceptibility to be vanishing in this phase. This
behavior occurs due to singularity of collinear states.  
Since all spin vectors in the {\it uuud} state are parallel 
to the magnetic field, spins can be rigid against the field. 
The {\it uuud} state hence
makes a plateau at the half magnetization $m/m_{\rm sat}=1/2$ 
in the magnetization process
(see Fig.~\ref{fig:mag_pro}). These results are obtained with 
the mean-field theory and also confirmed with Monte Carlo method. 
The {\it uuud} state can stably exist even at low but finite 
temperatures and it
also shows a finite-temperature phase transition due to the breakdown
of the translational symmetry. We will further discuss finite-temperature
properties in section V. 

\section{Quantum ($S=1/2$) model with $B=0$.}
\indent
When there are various degenerate ground states in the classical
limit, quantum effects play an essential role in forming the ground state
of the quantum model. One possibility is that a new quantum ground
state appears due to tunneling between the classical states. 
This mixture of
states can occur between the degenerate ground states in the same
$S^z$ sector. This possibility was
discussed as an origin of the disordered state which is observed in
low-density solid {}$^3$He films.\cite{KuboM,IshidaMYF} Another
possibility is that one of the degenerate ground states may be
selected due to quantum effects. 

\begin{figure}[tbhp]
  \begin{center}
    \leavevmode
    \epsfig{file=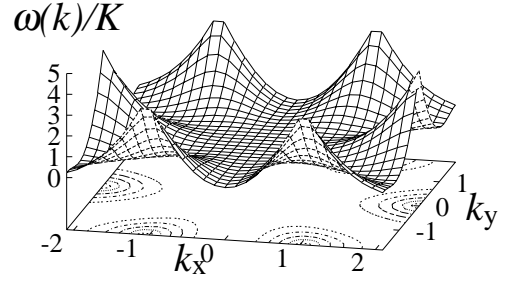,width=3.36in}
  \end{center}        
  \caption{The lowest mode of spin-wave spectrum on the {\it uuud}
    ordered state. The contours of constant energy are written on the 
    bottom plane. Zero
    modes appear on the three lines, $k_x=0$ and 
    $k_y = \pm \protect\sqrt3 k_x / 2$. }
  \label{fig:spectrum}
\end{figure}
To test these possibilities, we first examine the {\it uuud} state,
which is one of the classical ground states, 
with the spin-wave approximation. Using the Holstein-Primakoff
transformation, we expand the Hamiltonian up to the quadratic form of
bosons
\begin{equation}
  \label{eq:sw}
{\cal H} = -3NK + \sum_k \bfA_k^\dagger {\cal D}_k \bfA_k
\end{equation}
with $\bfA_k^\dagger =(a_k^\dagger,b_k^\dagger,c_k^\dagger,d_{-k})$
and 
\begin{equation}
  \label{matrix}
  {\cal D}_k = \left( 
\begin{array}{cccc}
      -4J & B_{\bfe_2,2\bfe_1} & 
      B_{\bfe_2-\bfe_1,2\bfe_1} & C_{\bfe_1} \\
      B_{\bfe_2,2\bfe_1} & -4J & 
      B_{\bfe_1,2\bfe_2} & C_{\bfe_1-\bfe_2} \\
      B_{\bfe_2-\bfe_1,2\bfe_1} & B_{\bfe_1,2\bfe_2} &
      -4J & C_{\bfe_2} \\
      C_{\bfe_1} & C_{\bfe_1-\bfe_2} & C_{\bfe_2} & 12(J+8K)
    \end{array} \right),
\end{equation}
where $a$, $b$, $c$ and $d$ denote
bosons for four sublattices and 
\begin{eqnarray}
    B_{\bfr,\bfr^\prime} &=& 4(J+2K)\cos \bfk\cdot \bfr 
    + 8K\cos \bfk\cdot (\bfr-\bfr^\prime),
\nonumber \\
    C_{\bfr} &=& 4(J+8K)\cos \bfk\cdot \bfr. 
\end{eqnarray}
Numerically diagonalizing this Hamiltonian with Copla's
method,\cite{Copla} we evaluate the excitation spectrum of spin
waves.\cite{KuboNM} The spectrum has four branches of excitation modes
in the 1st Brillouin zone of the four-sublattice structure. 
The lowest mode shows an ill behavior (see Fig.~\ref{fig:spectrum});
it has flat modes along three lines in momentum space, $k_x=0$ and 
$k_y = \pm \sqrt{3} k_x/2$, 
and furthermore the dispersion curves across the three
lines have a form $|\bfk-\bfk_0|^6$. These modes correspond to
the line excitations that we have shown for the classical model in 
Section II. 
This flat mode suggests that the {\it uuud} state does not have spin
stiffness. Higher-order terms of 
spin-wave expansions may destroy the {\it uuud} state due to
non-linear effects. 

We also studied the ground state of finite-size systems with 
exact-diagonalization method. The systems with the size
$N=12$, 16, 20, 24 and 28 are treated. (See Appendix A for shapes of
the finite-size clusters.) 
The results reveal that the ground state belongs to the $S=0$
space and hence it is not the {\it uuud} state for $-7K\le J$, which
almost covers the phase (b). 
Recent numerical study by Misguich et
al.\cite{MisguichBLW}\ also suggests that the ground state in the same
parameter region is spin liquid with $S=0$. 

All the above results are consistent with each other and give a
unique picture. The ground state belongs to the $S=0$ space and 
the {\it uuud} state has a little higher energy than the ground state 
due to quantum effects. The {\it uuud} state thus disappears
from the ground state in the quantum model. But it again becomes
stable under the magnetic field. We will discuss this point in the
next section. 

\section{Quantum ($S=1/2$) model under the magnetic field ($B>0$).}
\begin{figure}[bt]
  \begin{center}
    \leavevmode
    \epsfig{file=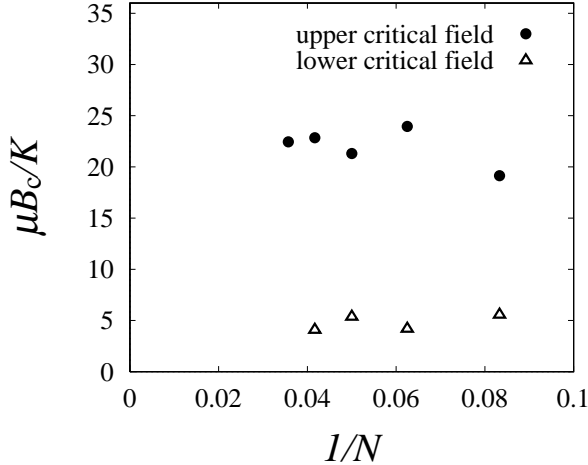,width=3.36in}
  \end{center}        
  \caption{Size dependence of the lower- and upper-critical fields of
    the magnetization plateau at $S^z_{\rm total}=N/4$ for the
    model with $J=-4$ and $K=1$.}  
  \label{fig:c_field}
\end{figure}

\begin{figure}[btp]
  \begin{center}
    \leavevmode
    \epsfig{file=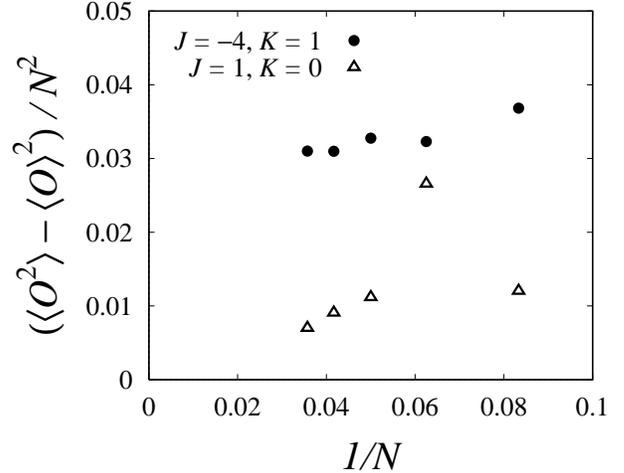,width=3.36in}
  \end{center}        
  \caption{Size dependence of long-range order of the {\it uuud}
    structure, 
    $(\langle {\cal O}^2 \rangle - \langle {\cal O}\rangle^2)/N^2$,
    in the ground state of the $S^z_{\rm total}=N/4$ space. Black
    circles denote data for the model with 
    $J=-4$ and $K=1$. Data for the Heisenberg antiferromagnet ($J=1$, 
    $K=0$) are also shown with triangle symbols for a comparison.} 
  \label{fig:uuud_order}
\end{figure}

\indent
To test the appearance of magnetization plateau at $m/m_{\rm sat}=1/2$
which we found in the classical limit, we investigate ground states of 
the quantum model (\ref{Hamiltonian}) in finite-size systems under the 
magnetic field. We study finite-size
($N\le28$) systems with periodic-boundary conditions. 
Figure~\ref{fig:mag_pro} shows the magnetization process of
finite-size systems with $J=-4$ and $K=1$. Though the magnetization
increases  
stepwisely due to finite-size effects, there clearly exists a broad
plateau at $m/m_{\rm sat}=1/2$ in every-size data. 
Width between the lower and
upper critical fields of the plateau does not vanish, as the system
size increases, and remains to be significantly large (see 
Fig.~\ref{fig:c_field}). This result strongly suggests that this
magnetization plateau survives in the thermodynamic limit. To examine 
whether this $m/m_{\rm sat}=1/2$ state has the {\it uuud}
long-range order, we consider the following {\it uuud} order parameter
\begin{equation}
  \label{o-param}
  {\cal O}=\dfrac{1}{2}\biggl( \sum_{i\in A} \sigma_i^z +\sum_{i\in B}
  \sigma_i^z  
  +\sum_{i\in C} \sigma_i^z -\sum_{i\in D} \sigma_i^z  \biggr)
\end{equation}
and calculate long-range order 
\begin{equation}\label{LRO}
( \langle {\cal O}^2 \rangle - \langle {\cal O} \rangle^2 )/N^2 
\end{equation}
in the ground state of the $S^z_{\rm total}=N/4$ space. 
Results are shown 
in Fig.~\ref{fig:uuud_order}, which clearly suggest that data are
extrapolated to a finite value in the $N\rightarrow\infty$ limit for
$J=-4K$. The extrapolated value is estimated as about 0.03 in the
$N\rightarrow\infty$ limit. Long-range order of the {\it uuud}
structure brings breakdown of
translational symmetry in the thermodynamic limit. A careful
estimation (see Appendix~B) leads that, when the translational
symmetry is spontaneous broken, the expectation value of spin on each 
sublattice is $\langle \sigma_i^z/2 \rangle=0.45$ for $i\in A$-, $B$-
or $C$-sublattice, and $\langle \sigma_i^z/2 \rangle=-0.35$ for 
$i\in D$-sublattice, respectively.  
The ground state in the $S^z_{\rm total}=N/4$ space thus has a rigid 
{\it uuud} long-range order and deviation of the sublattice
magnetization from the classical value is 
small. By the way, the total magnetization in the {\it uuud} ordered
state does not change from the classical 
value $m/m_{\rm sat}=1/2$ by quantum effects. 
Note that the expectation values of spin satisfy 
$\sum_i \langle \sigma_i^z/2 \rangle/N = (3 \times 0.45 - 0.35)/4 =1/4$.
The spin-wave analysis also does not give any quantum correction
to the total magnetization in the {\it uuud} state. 

Magnetization
plateaus of two-dimensional systems have been observed both
theoretically and experimentally. On the triangular lattice, a
magnetization plateau was observed at $m/m_{\rm sat}=1/3$ in the
measurements of C$_6$Eu\cite{SuematsuOSSMD} and
CsCuCl$_3$,\cite{NojiriTM} and in theoretical studies of
antiferromagnets (with two-spin exchange).\cite{Kawamura,ChubukovG,Nikuni} 
The plateau comes from the three-sublattice {\it uud} ordered state. 
The quantum correction to the magnetization
$m/m_{\rm sat}=1/3$ is also vanishing.\cite{ChubukovG,Nikuni} 
For MSE models, a similar magnetic plateau
was also observed at $m/m_{\rm sat}=1/2$ by finite-size studies. 
Roger and Hetherington first discovered the magnetic plateau in a MSE
model with 
four-spin exchange on the square lattice.\cite{RogerH,mean-field} 
(In the mean-field approximation, the ground state of this plateau is
also the {\it uuud} state. (See Ref.\ \onlinecite{mean-field}))
Recently Misguich et al.\ also observed a plateau in a model
with two-, four-, five- and six-spin exchanges on the triangular
lattice. We believe that these
plateaus in MSE models come from the {\it uuud} ordered state and that
the four-spin exchange is the main cause for the plateau as well. 

\begin{figure}[tbp]
  \begin{center}
    \leavevmode
    \epsfig{file=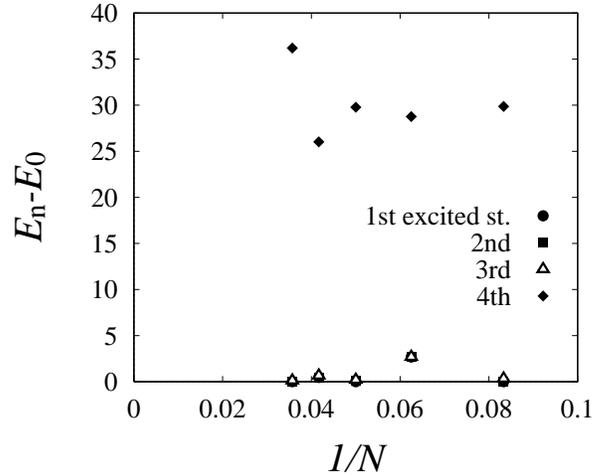,width=3.36in}
  \end{center}
  \caption{Size dependence of excitation energy of four low-lying
    excited states in the 
    $S^z_{\rm total}=N/4$ space for the model with $J=-4$ and
    $K=1$. The 
    symbol for the third excited state almost overlaps those for the 
    first and second states.} 
  \label{fig:excitation}
\end{figure}

Next, we study the excitations in the $S^z_{\rm total}=N/4$ space. 
Figure~\ref{fig:excitation} shows excitation energy of up to fourth 
excited state. Three low-lying excited states have energy very close
to the ground-state one and they converge to the ground
state as the system size is enlarged. Above them there is a large gap, 
which seems not to vanish in the $N\rightarrow\infty$ limit. Thus four 
states exist around the lowest level and are clearly separated from
other excited states. The
number of low-lying levels, four, is equal to the degeneracy of 
{\it uuud} states, which comes from the spatial translation of
sublattices. Furthermore, the four
low-lying states have the same translation group as the following
schematic states, respectively, 
\begin{eqnarray}
  \label{four-state}
  |1\rangle &=& (1+{\cal R}_1+{\cal R}_2+{\cal R}_1 {\cal R}_2)
               |\uparrow\uparrow\uparrow\downarrow\rangle, \nonumber\\
  |2\rangle &=& (1+{\cal R}_1-{\cal R}_2-{\cal R}_1 {\cal R}_2)
               |\uparrow\uparrow\uparrow\downarrow\rangle, \nonumber\\
  |3\rangle &=& (1-{\cal R}_1+{\cal R}_2-{\cal R}_1 {\cal R}_2)
               |\uparrow\uparrow\uparrow\downarrow\rangle, \nonumber\\
  |4\rangle &=& (1-{\cal R}_1-{\cal R}_2+{\cal R}_1 {\cal R}_2)
               |\uparrow\uparrow\uparrow\downarrow\rangle, 
\end{eqnarray}
where $|\uparrow\uparrow\uparrow\downarrow\rangle$ denotes one of the
{\it uuud} states, $\otimes_{i\in A,B,C} |\uparrow\rangle_i  
\otimes_{j\in D} |\downarrow\rangle_j$, and ${\cal R}_i$ ($i=1$, 2) mean 
the translation of sites by unit vectors $\bfe_i$. We hence believe that
these four levels form four ground states which have {\it uuud} order 
in the thermodynamic limit and translational invariance is
spontaneously broken in the ground states. 
This argument leads to a conclusion that the
fifth low-lying state corresponds to the lowest excitation in the
thermodynamic limit and hence the spin excitation spectrum has a
finite gap in the same $S^z_{\rm total}$ sector. 

The appearance of the magnetization plateau and the excitation gap can 
be understood by introducing a particle picture into the spin
model.\cite{Totsuka} 
Let us regard the down spins as particles moving
in background of up spins and the magnetic field $\mu B$ as minus of the
chemical potential of the particle.\cite{MatsubaraM} 
Then the present spin system is 
mapped to a hard-core boson system and we can recognize the {\it uuud} 
ordered state as a Mott insulating state with charge-density
wave (CDW). Note that this density
wave can be ordered by repulsion between particles on 
nearest- or next-nearest-neighbor pairs of sites, which 
originates from the four-spin exchange interaction. 
In incompressible CDW, particles are insulating and charge-density
excitations have a finite gap. And the compressibility is
vanishing. Through the mapping, 
these features correspond to the finite excitation gap and the magnetization 
plateau of the original spin model. Thus the
magnetization plateau can be understood as an insulating-conducting
transition in the particle picture. 
We hence have an explanation for the plateau from the
particle limit ($S=1/2$).\cite{Totsuka2} 
As we mentioned in section II, the
appearance of the plateau in the $S\rightarrow\infty$ limit originates
from the rigidity of the collinear state. 
If the scenario in the particle picture is valid, the mechanism for
the appearance of plateau is understood both from the classical limit 
($S\rightarrow\infty$) and the particle limit ($S=1/2$). Note that, in
both cases, the {\it uuud} order is the key property for the appearance of 
the magnetization plateau. 

The 1/2-plateau appears for a wide parameter region. In
Fig.~\ref{fig:PD}, we show the phase boundary of the parameter region
of the plateau. Except for weak magnetic field cases, the region of
the $m/m_{\rm sat}=1/2$ phase 
becomes wider in comparison with the classical model. Quantum
effects thus stabilize the plateau and enhance its appearance. 

\section{Thermal effects}
\indent
It is also important to discuss thermal effects to the magnetization
plateau for a comparison with experiments at finite temperatures. 
A favorable property of the {\it uuud} order is that it 
accompanies a phase transition at a finite temperature. 
Since the {\it uuud} order is a discrete symmetry breaking, i.e., 
the number of ground states is four under the magnetic field, the
symmetry breakdown occurs even at low but finite temperatures in two
dimensions. The {\it uuud} order hence survives against thermal
fluctuations and a phase transition occurs at a finite temperature. 
(Note that this argument does not depend on whether the
system is quantum or classical.) 
Fortunately, the magnetization plateau survives even at low 
temperatures due to ordering of the {\it uuud} structure. 
Magnetization in the {\it uuud} ordered state is 
close to $m/m_{\rm sat}=1/2$ at low temperatures and it makes a
plateau (or a shoulder) near the half of the saturated magnetization. 
This argument comes from only the degeneracy of the ground states, 
using the aspect of universality in phase transition. 

To confirm the above argument, we study the finite-temperature 
properties in the classical limit using Monte Carlo simulations. 
We believe that critical properties of the phase transition at finite 
temperatures are governed by thermal fluctuations and quantum effects
do not change its universality, though the value of
the critical temperature or order parameter will
deviate from the quantum one. 
Monte Carlo simulations were performed with 
the Metropolis algorithm. If a spin flip is rejected, we randomly 
rotate the spin about the local molecular field. 
We construct finite-size systems with a unit cluster which has 12 sites. 
The system sizes are $N=12 L^2$ with $L=4$, 6, 8, 12 and 
16 with periodic-boundary conditions. 
After discarding initial 30000--50000 Monte Carlo steps per spin (MCS) 
for equilibration, subsequent $3\times 10^5$--$5\times 10^5$ MCS are 
used to calculate the average.

\begin{figure}[tbp]
  \begin{center}
    \leavevmode
    \epsfig{file=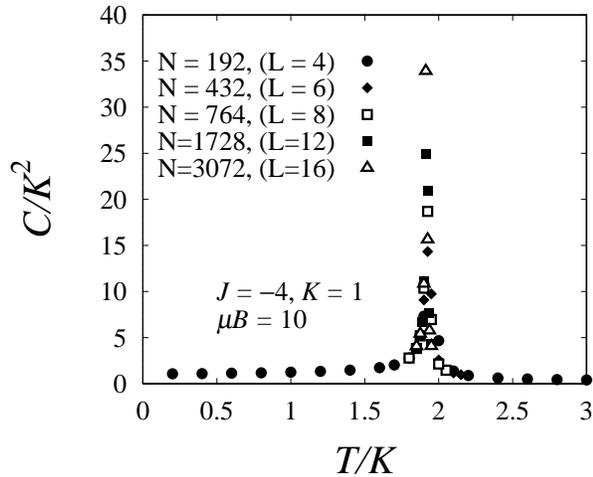,width=3.36in}
  \end{center}    
  \caption{Temperature dependence of the specific heat of the
    classical model with $J=-4$, $K=1$ and $\mu B=10$. Monte Carlo
    simulations were done on triangular lattices with finite size $N=12 
    L^2$. } 
  \label{fig:s_heat}
\end{figure}
\begin{figure}[tbp]
  \begin{center}
    \leavevmode
    \epsfig{file=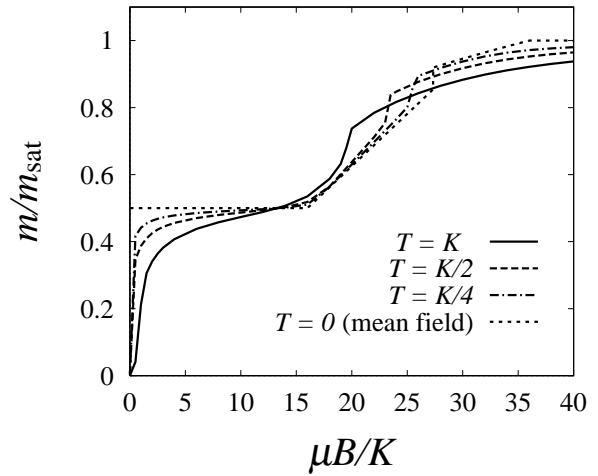,width=3.36in}
  \end{center}    
  \caption{Magnetization process of the classical model with $J=-4$
    and $K=1$ at finite temperatures $T/K=0.25$, 0.5 and 1.0. The
    magnetization process at zero temperature, which we obtained
    with the mean-field theory,\protect\cite{KuboM} is also shown for
    a comparison.}  
  \label{fig:mag_pro_finite_T}
\end{figure}

Monte Carlo simulations were done for the model under a 
magnetic field which realizes the {\it uuud} ground state. Figure 
\ref{fig:s_heat} shows specific-heat data for the case $J=-4$, $K=1$
and $\mu B=10$. The data show a sharp divergence around $T=1.9K$. We
also found that the {\it uuud} long-range order exists below the
critical temperature. 
Near the critical temperature, the
magnetization curve is still rounded and smooth. As lowering the
temperature, the slope of the curve decreases and becomes flat around 
$m/m_{\rm sat}=1/2$, as shown in fig.\ \ref{fig:mag_pro_finite_T}. 
Misguich et al.\cite{MisguichBLW}\ also demonstrated stability of the 
plateau at finite low temperatures in the quantum MSE model on a
finite-size ($N=16$) lattice. 
Similarly, in the Heisenberg antiferromagnet on the triangular
lattice, the magnetization plateau at $m/m_{\rm sat}=1/3$ was
successfully observed at finite temperatures in the measurements of
C$_6$Eu\cite{SuematsuOSSMD} and CsCuCl$_3$,\cite{NojiriTM} and 
in Monte Carlo simulations.\cite{Kawamura} 

The critical temperature is estimated as $T_{\rm c}=1.9K$ for the
model with $J=-4K(<0)$ and $\mu B=10K$. For a weak magnetic field $\mu 
B=5K$, the estimate of $T_{\rm c}$ is about $1.7K$. These values can
change due to quantum effects in the $S=1/2$ model. 
From the degeneracy of the ground states, the phase transition is 
expected to be of second order and to belong to the 4-state Potts
universality class. 
The finite-size scaling analysis, however, shows that critical
exponent $\alpha$ is much larger than the expected value
$\alpha=2/3$ for the 4-state Potts model.\cite{Wu} Moreover
distribution of energy histogram in Monte Carlo simulations has two
peaks at the critical temperature, which suggests that 
the phase transition is of first order. A similar deviation of the
universality was also found in the phase transition
of chiral symmetry breaking in the model (\ref{Hamiltonian}) for strong 
$K$.\cite{MomoiKN,MomoiKN2}
These deviations may come from frustration effects, or singularity of 
four-body interactions. Critical properties of these
phase transitions will be discussed further in a forthcoming paper.

The discrete symmetry of the {\it uuud} order in a magnetic field 
accompanies two favorable properties 
that a sharp phase transition occurs at a finite temperature and 
the magnetization plateau stably exists at low but finite 
temperatures. We expect that these properties make the experimental
observation of the magnetization plateau possible. 

\section{Summary and discussion}
\indent
In this paper, we examined the appearance of the magnetization plateau at
$m/m_{\rm sat}=1/2$ in a 2D MSE model on the triangular
lattice, which we predicted in our previous work. 
This plateau appears when the two- and four-spin exchange
interactions compete strongly, which may be realized in 2D low-density 
solid {}$^3$He, and if the five- and six-exchanges are not too strong. 
The four-spin exchange is important and
relevant to make the plateau at $m/m_{\rm sat}=1/2$. 
Since ordering of {\it uuud} structure is breakdown of discrete
symmetry, it accompanies a phase transition at a finite temperature. 
Then the plateau would appear below the critical temperature. 
Experimental observation of this
plateau in the magnetization process of the solid {}$^3$He films will
confirm that the four-spin exchange is strong and important. 

Finally we discuss possibility of observing plateau in
experiments. 
Comparing our results with measurements of susceptibility and
specific heat, we estimate the effective exchanges $J$ and $K$ for
low-density films. Susceptibility measurements at low densities 
show that the effective coupling $J_{\chi}$ is antiferromagnetic,
$J_{\chi}>0$, though it is close to vanishing.\cite{RogerBBCG} 
In our model (\ref{Hamiltonian}), the effective coupling behaves as 
$J_{\chi}=2(J+6K)$ and hence we can estimate as $J\gtrsim-6K$. 
Specific heat measurements indicate that energy (spin) gap is small or 
vanishing,\cite{IshidaMYF} where the system might be close
to the critical
point between the ferromagnetic phase and the liquid
phase.\cite{MisguichBLW} 
Since the critical point is around $J/K\simeq-8$ in the model
(\ref{Hamiltonian}), the exchange parameters can be estimated as 
$J\gtrsim-8K$ from the specific-heat data. 
The value $J/K$ seems to be close to the boundary of the region where
1/2-plateau appears (see Fig.~\ref{fig:PD}), though two estimates are
not consistent with each other. In real systems, more-than-four-spin
exchanges may move the boundary to right or left (in
Fig.~\ref{fig:PD}). For quantitative arguments, further studies will be
needed on the effect of five- or six-spin exchanges. To make 
the plateau stable and wider, we would need lower-density solid films and
enlarge the four-spin exchange effect. It was reported that, by
preplating HD on grafoil, {}$^3$He atoms solidify at lower density than 
double layer {}$^3$He films.\cite{SiqueiraLCS93,SiqueiraNCSb} 
Thus solid {}$^3$He preplated HD may be a
plausible candidate for observing the magnetization plateau. 
In our 
model with $J=-4K(<0)$, the lower critical field of the plateau is
estimated about $4K/\mu$. 
Setting the parameter as $K=1.0$(mK) and $\mu=2.13\mu_{\rm N}$, 
where $\mu_{\rm N} = 0.366$(mK/T), we estimate the lower
critical field as $B_c \simeq 5 [T]$. (The value may be changed by
other multiple-spin exchanges, i.e.\ six-spin exchange, in real
systems.) Since this magnitude of the
field is accessible with the present experimental equipments, we
expect experimental 
verification of the magnetization process to be possible in 2D
low-density solid {}$^3$He.

\acknowledgements
The authors would like to thank Hiroshi Fukuyama and Hikaru Kawamura 
for stimulating discussions and comments. 
One of the authors (T.M.) acknowledges Keisuke Totsuka for useful 
discussions and K.K.\ acknowledges J.\ Saunders, C.\ Lhuillier and 
G.\ Misguich for useful discussions. 
This work was supported by Grant-in-Aid Nos.\ 09440130 and 10203202 
from the Ministry of 
Education, Science and Culture of Japan. 
The numerical calculations were done on 
Facom VPP500 at the ISSP of the University of Tokyo and DEC Alpha 500
personal workstation at the Institute of Physics of the University of
Tsukuba. K.K.\ also acknowledges the EPSRC Visiting Fellowship Grant
GR/L 90804 and the kind hospitality at the Department of Mathematics,
Imperial College, London, where a part of this work  was
accomplished. 

\appendix
\section{Clusters used in finite-size studies.}
\begin{figure}[h]
  \begin{center}
    \leavevmode
    \epsfig{file=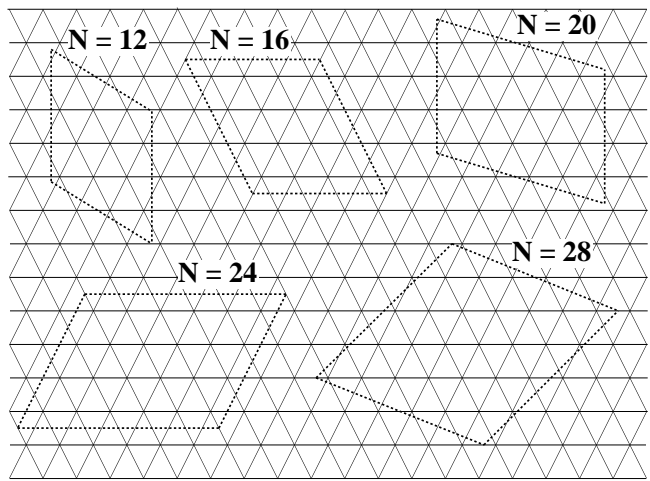,width=3.36in}
  \end{center}    
  \caption{Finite-size clusters which we used in the exact
    diagonalization study.}
  \label{fig:cluster}
\end{figure}

In the numerical study of the $S=1/2$ model on finite-size systems, 
we used finite-size
clusters with the periodic boundary condition which matches the
four-sublattice structure. Clusters are set on the
triangular lattice as shown in Fig.~\ref{fig:cluster}. 

\section{Evaluation of expectation values of the sublattice magnetization.}
Estimation of the sublattice magnetization from the long-range order 
(\ref{LRO}) needs a careful consideration about symmetry breaking. In
the thermodynamic limit, translational symmetry is spontaneously
broken in a pure natural ground state with {\it uuud} order. But the
ground states of finite-size 
systems become mixed symmetric ones because of finite-size effects. 

Here we write the mixed symmetric ground state in the $S^z_{\rm
  total}=N/4$ space as $|\phi\rangle$. We also consider a state
$|\psi_1\rangle$ in
which translational symmetry is broken and whose thermodynamic limit
is the natural pure ground state. 
By the space translation of the state, $|\psi_1\rangle$ relates to 
other three {\it uuud} ordered states in the form 
\begin{equation}
|\psi_2\rangle = {\cal R}_1 |\psi_1\rangle,\ \ 
|\psi_3\rangle = {\cal R}_2 |\psi_1\rangle,\ \ 
|\psi_4\rangle = {\cal R}_1 {\cal R}_2 |\psi_1\rangle,
\end{equation}
where ${\cal R}_1$ (${\cal R}_2$) denotes spatial translation by $\bfe_1$
($\bfe_2$). 
In the mixed state, there is translational
invariance, 
\begin{equation}
\langle \phi | S_i^z | \phi \rangle 
=\dfrac{1}{4} \mbox{\hspace*{1cm}for any site $i$}.
\end{equation}
On the other hand, in the {\it uuud} ordered state $|\psi_1\rangle$, 
the translational symmetry is broken, 
\begin{eqnarray}
\langle \psi_1 | S_i^z | \psi_1 \rangle &=& 
m_1 \mbox{\hspace*{1cm}for $i\in A,B$ or $C$,}\\
\langle \psi_1 | S_i^z | \psi_1 \rangle &=& 
m_2 \mbox{\hspace*{1cm}for $i\in D$},
\end{eqnarray}
where $m_1\ne m_2$. Since the total magnetization is
$N/4$, the sublattice magnetization $m_1$ and $m_2$ satisfy 
\begin{equation}\label{eq:magnetization}
\dfrac{3}{4} m_1 + \dfrac{1}{4} m_2 = \dfrac{1}{4}.
\end{equation}
It is expected that the symmetric mixed state can be decomposed into
the pure states in the thermodynamic limit in the form 
$|\phi\rangle=(|\psi_1\rangle + e^{i\theta}|\psi_2\rangle 
+ e^{i\psi}|\psi_3\rangle + e^{i\phi}|\psi_4\rangle )/2$,
where $\theta$, $\psi$ and $\phi$ denote arbitrary real numbers. 
The following relation hence holds in the large $N$ limit: 
\begin{eqnarray}\label{eq:relation}
\lefteqn{\dfrac{1}{N^2} \langle \phi | {\cal O}^2 |\phi \rangle
=\dfrac{1}{4 N^2} ( \langle \psi_1 | {\cal O}^2 |\psi_1 \rangle}\\
& &\ \ \ + \langle \psi_2 | {\cal O}^2 |\psi_2 \rangle
+ \langle \psi_3 | {\cal O}^2 |\psi_3 \rangle
+ \langle \psi_4 | {\cal O}^2 |\psi_4 \rangle ). \nonumber
\end{eqnarray}
From the numerical calculations in section IV, we have 
\begin{equation}
\lim_{N\rightarrow \infty}
\dfrac{1}{N^2} \langle \phi | {\cal O}^2 |\phi \rangle
=0.03+\biggl(\dfrac{1}{8}\biggr)^2.
\end{equation}
For the pure state, the clustering property of the state leads 
\begin{eqnarray}
\lim_{N\rightarrow \infty}
\dfrac{1}{N^2} \langle \psi_1 | {\cal O}^2 | \psi_1 \rangle
&=&\biggl(\lim_{N\rightarrow \infty} \dfrac{1}{N} \langle \psi_1 |
{\cal O} |\psi_1\rangle\biggr)^2 \nonumber\\
&=&\biggl\{\dfrac{1}{4}(3m_1 - m_2)\biggr\}^2
\end{eqnarray}
and
\begin{eqnarray}
\lim_{N\rightarrow \infty}
\dfrac{1}{N^2} \langle \psi_2 | {\cal O}^2 | \psi_2 \rangle
&=& \biggl(\lim_{N\rightarrow \infty} \dfrac{1}{N} \langle \psi_2 | 
{\cal O}|\psi_2 \rangle\biggr)^2 \nonumber\\
&=& \biggl\{\dfrac{1}{4}(m_1 + m_2)\biggr\}^2. 
\end{eqnarray}
In the same way, 
\begin{eqnarray}
\lim_{N\rightarrow \infty}
\dfrac{1}{N^2} \langle \psi_3 | {\cal O}^2 | \psi_3 \rangle
&=& \lim_{N\rightarrow \infty}
\dfrac{1}{N^2} \langle \psi_4 | {\cal O}^2 | \psi_4 \rangle\nonumber\\
&=& \biggl\{\dfrac{1}{4}(m_1 + m_2)\biggr\}^2. 
\end{eqnarray}
Inserting these relations into eq.~(\ref{eq:relation}) and using
eq.~(\ref{eq:magnetization}), we
have 
\begin{equation}
48{m_1}^2-24m_1+1.08=0
\end{equation}
and then we obtain two solutions
$(m_1,m_2)=(0.45,-0.35)$ and $(0.005,0.985)$. From the constraints
$|m_1|\le0.5$ and $|m_2|\le0.5$, we conclude that the
former one is the physical solution.



\begin{thebibliography}{99}
\bibitem{MomoiKN2} Critical exponents were estimated as $\alpha=0.62\pm 
  0.05$ and $\beta=0.090\pm 0.008$ from finite-size scaling analysis. 
  However there remains another possibility that the phase transition
  is of first order and critical exponents are wrongly estimated
  due to smallness of system size. Indeed, distribution of energy
  histogram for 
  systems with large size ($N\ge 3072$) have two peaks which are very
  close to each other near the critical temperature.
\end{thebibliography}
\end{document}